\documentclass[preprint,prd,tightenlines,nofootinbib]{revtex4}
\usepackage{amsmath}
\usepackage{amsxtra}
\usepackage{graphicx}

\def\lmunu{L^{\mu\nu}}

\def\kmu{k^\mu}
\def\knu{k^\nu}
\def\qmu{q^\mu}
\def\qnu{q^\nu}
\def\gmunu{g^{\mu\nu}}

\def\mxx{m_\chi^2}

\def\edm{\mathcal{D}}

\begin{document}

\title{Detecting Dipolar Dark Matter in Beam Dump Experiments}
\author{Subhendra Mohanty}
\affiliation{
Physical Research Laboratory,\\
Navarangapura, Ahmedabad 380009, India.
}
\author{Soumya Rao}
\affiliation{
	ARC Centre of Excellence for Particle Physics at the Terascale,\\
	Department of Physics, University of Adelaide, Adelaide, SA 5005, Australia.
}
\begin{abstract}
	We study interaction of low mass dark matter within beam dump experiments.  In
	particular we study the dipolar dark matter model which assumes that the dark
	matter couples to Standard Model particles via its electric or magnetic dipole
	moment.  We analyse the constraints on this model in the context of a particular
	beam dump experiment E613 conducted in the Fermilab.  We find that dark matter
	mass in the range of $1-10$ GeV with a magnetic dipole moment between
	$(0.33-1.5)\times 10^{-7}\mu_B$ and a electric dipole moment between
	$(0.5-3)\times 10^{-17}$ e-cm.  We compare the bounds from other experimental
	data, such as helioseismological data and direct detection experiments.
\end{abstract}
\maketitle

\section{Introduction} 

Evidence from galactic rotation curves as well as bullet cluster collisions have strongly
suggested the existence of dark matter(DM).  However the nature of DM is unknown.  In
theory there exist many candidates for DM, the most popular of them being the case of
weakly interacting massive particle (WIMP).  The WIMP scenario is realized in many models,
with neutralino in supersymmetric theories being the most well studied of them.  However,
there also exists the possibility of DM interacting electromagnetically with Standard
Model (SM) particles\cite{bagnasco, pospelov, sigurdson, gardner, masso, cho, chang,
barger, zurek, gondolo}.  We refer to this model of DM interactions as dipolar DM model.

The dipolar DM model assumes that dark matter couples to photons through loops to give
rise to electric and magnetic dipole moments.  
Here we study the case in which DM particles are Dirac fermions.  The effective
Lagrangian for such a DM particle interacting with an electromagnetic field
($\mathcal{F}{\mu\nu}$) through its electric dipole moment ($\edm$) and magnetic dipole
moment ($\mu$) is

\begin{equation}
	\mathcal{L}_{ddm}=-\frac{i}{2}\bar{\chi}\sigma_{\mu\nu}(\mu+\gamma^5\edm)\mathcal{F}^{\mu\nu}
	\label{lag}
\end{equation}

Here we consider this particular scenario in the context of DM detection in so called beam
dump experiments also known as fixed target experiments.  Complimenting the searches from
direct and indirect detection experiments, beam dump experiments feature a high energy
beam incident on a fixed target.  The DM particles produced from this collision are then
detected in a suitable detector.  In this work we particularly focus on the E613
experiment at Fermilab in which a 400 GeV proton beam is incident on a tungsten target
with the resultant DM produced from the annihilation of beam proton and target proton
being detected in a lead detector after passing through iron shielding.  The advantage
over direct detection in such experiments is that of higher luminosity.  But the reach of
such experiments and in particular of E613 is restricted to low mass DM due to the
kinematics of fixed target experiments.  Future experiments like the new fixed target
facility proposed at the CERN SPS called SHiP (Search for Hidden Particles) \cite{ship}
can explore this possibility.  The possibility of DM detection at beam dump experiments
has been studied in the case of light mediators in the dark sector mediating DM
interactions with SM particles \cite{batell,essig} and also in the case of $Z^\prime$ as
the mediating particle \cite{dobrescu}.  More recently a similar model with dark vector
boson was studied in the context of E613 \cite{tait}.  We follow the approach in
\cite{tait} and study the constraints on the dipolar DM model from E613.

The paper is organized as follows :  In section 2 we describe the method for calculating
the production cross section of dipolar DM when the 400 GeV proton beam strikes the
tungsten target.  In section 3 we describe the deep inelastic scattering that takes place
between the DM and the lead target nuclei.  In section 4 we give the results from
analysing the dipolar DM in the case of E613 and compare constraints from other
experiments.  Finally we conclude in section 5.

\section{Dipolar DM production in E613}

Here we follow the procedure outlined in \cite{tait}.  In a beam dump experiment like
E613, DM particles are produced through t-channel annihilation process from constituent
quarks of protons in the beam and the target nucleus.  In case of the E613 we have a 400
GeV proton beam striking a tungsten target.  The DM particles are produced from the
process, $pp\to \bar{\chi}\chi+X$.  The cross section for this hard process is calculated
by incorporating the Lagrangian in eq.(\ref{lag}) into MADGRAPH 5 \cite{madgraph} using
FEYNRULES \cite{feynrules}.  The number of $\chi$'s produced is then given by
\begin{equation}
	\frac{dN}{dEd\theta}=n_t~N_{beam}~L_t~\frac{d\sigma(pp\to\chi\chi)}{dEd\theta}
	\label{dnde}
\end{equation}
where $n_t$ is the number density of nucleons inside the tungsten target, $N_{beam}$ is
the number of protons in the beam that are incident on the tungsten target and $L_t$ is
the length of the tungsten target.  The geometry of the E613 detector is such that only
those DM particles are accepted for which the scattering angle $\theta < 0.0134$
\cite{tait}.  This is a conservative limit on the scattering angle compared to the
detector acceptance in the original experiment where $\theta < 0.037$.  Thus we integrate
eq.(\ref{dnde}) for $0<\theta<0.0134$ to obtain DM distribution $dN/dE$.

\section{Deep Inelastic scattering of Dipolar DM}

Following \cite{tait} we place an energy cut of 20 GeV as the minimum energy required to
register an event in the detector.  Now in order to find the number of events we need to
find scattering cross section of the DM produced in the experiment with the detector
nuclei, having energy $E_\chi$ where $E_\chi>20$ GeV.  Since these DM particles are
produced in a collision with CM energy of $400$ GeV, their scattering off the lead nuclei
in the detector is deeply inelastic.  The energetic DM particle therefore undergoes a deep
inelastic scattering with the lead nucleus inside the detector via photon exchange.  The
DM couples to the photon through its electric/magnetic dipole moment.  The initial DM
momentum before scattering being $k$ while that after scattering being $k^\prime$ the
momentum transfer carried by the photon is $q=k-k^\prime$.  Now using the formalism of
deep inelastic scattering of leptons we define the Bjorken scaling variable
$x=\dfrac{Q^2}{2m_N \nu}$, with $Q^2=-q^2$ and $\nu$ being the energy of the photon in the
rest frame of the nucleus.  With this the differential scattering cross section in terms
of leptonic matrix element $\lmunu$ and hadronic matrix element $W_{\mu\nu}$ is given by

\begin{equation}
	\frac{d\sigma}{d\nu dQ^2}=\frac{e^2 g_{dipole}^2}{16\pi m_N(E_\chi^2-m_\chi^2)}
	\frac{\lmunu W_{\mu\nu}}{Q^4}
	\label{dsc}
\end{equation}
where $g_{dipole}=\mu$, with $\mu$ being the magnetic dipole moment of DM and for DM
interacting via the electric dipole moment, $g_{dipole}=\mathcal{D}$.

Now the leptonic current for the magnetic dipole moment interaction of DM corresponding to
the Lagrangian in eq.(\ref{lag}) is
\begin{equation}
	\lmunu = Q^2\left[4\kmu\knu-2\left(\kmu\qnu+\qmu\knu \right)+\qmu\qnu\right]
	-4\mxx\left(Q^2\gmunu+\qmu\qnu\right)
	\label{lmdm}
\end{equation}
And similarily for the electric dipole moment interaction we have
\begin{equation}
	\lmunu = Q^2\left[4\kmu\knu-2\left(\kmu\qnu+\qmu\knu \right)+\qmu\qnu\right]
	\label{ledm}
\end{equation}

The hadronic matrix element $W_{\mu\nu}$ can be written in terms of structure functions
such that we separate out contributions from longitudinally polarized photons and
transversely polarized photons as \cite{tait}

\begin{equation}
	W_{\mu\nu}=\left( -g_{\mu\nu}+\frac{q_\mu q^\nu}{q^2}+2xa_{\mu\nu}
	\right)F_T(x,Q^2)+a_{\mu\nu}F_L(x,Q^2)
	\label{wmunu}
\end{equation}
where
\begin{equation}
	a_{\mu\nu}=\frac{1}{p\cdot q+2xm_N^2}\left( p_\mu-\frac{p\cdot q}{q^2}q_\mu
	\right)\left( p_\mu-\frac{p\cdot q}{q^2}q_\mu \right)
	\label{amunu}
\end{equation}

At the lowest order in perturbation theory the structure function $F_L=0$ while
$F_T=\dfrac{1}{2x}\sum\limits_q xf(x,Q^2)$.

Contracting the leptonic and hadronic tensors in eqs. (\ref{lmdm}) and (\ref{wmunu}) we
have for the magnetic dipole interaction

\begin{equation}
	d\sigma=\dfrac{e^2\mu^2}{16\pi}\dfrac{d\nu dQ^2}{E^2-\mxx}\dfrac{\nu}{Q^4}\left[
	\dfrac{Q^2\left( 2E -\nu \right)^2}{\nu^2 +Q^2} -Q^2 +4\mxx\right]\sum\limits_q
	xf_{q/A}(x,Q^2)
	\label{eq21}
\end{equation}

Similarily for the electric dipole interaction we have
\begin{equation}
	d\sigma=\dfrac{e^2\edm^2}{16\pi}\dfrac{d\nu dQ^2}{E^2-\mxx}\dfrac{\nu}{Q^4}\left[
	\dfrac{Q^2\left(2E-\nu\right)^2}{\nu^2+Q^2}-Q^2-4\mxx\right]\sum\limits_q
	xf_{q/A}(x,Q^2)
	\label{ed21}
\end{equation}

For the nuclear parton distribution functions we use those provided by Hirai
etal. \cite{hirai}.  The above expression when integrated over $\nu$ and $Q^2$ gives the
cross section for scattering of DM with nucleon inside the target.  The limits of
integration are as follows:

\begin{align}
	E_{\mathrm{cut}}&<\nu<E-m_\chi \\
	Q_l^2&<Q^2<4(k^2-E\nu)-Q_l^2
	\label{lim}
\end{align}
where $Q_l^2=\dfrac{2m_\chi^2\nu^2}{k^2-E\nu+\sqrt{(k^2-E\nu)^2-m_\chi^2\nu^2}}$
with $k^2=E^2-m_\chi^2$.

In addition to the above limits one also has the upper limit $x<1$ which translates to
$Q^2<2m_N\nu$.
From the cross section so obtained we can now write the mean free path of the DM
particle as

\begin{equation}
	\lambda=\frac{1}{\rho\sigma(\chi N\to\chi N)}
	\label{mfp}
\end{equation}

where $\rho$ is the number density of the nucleon inside the target material and $\sigma$
is the scattering cross section of DM from a nucleon.  The probability of a DM particle
scattering inside the detector is then given by $P=1-e^{-L/\lambda}$ and for a DM particle
that behaves like a WIMP interacting through a weak dipole moment we can use the
approximation $P\sim L/\lambda$.

Finally the number of expected events in the detector is
\begin{equation}
	N_\mathrm{ev}=\int dE \left[P_\mathrm{Pb}(1-P_\mathrm{Fe})\right]\frac{dN}{dE}
	\label{ndet}
\end{equation}
where $P_\mathrm{Pb}$ is the probability of DM scattering inside the lead detector and
$P_\mathrm{Fe}$ is that inside the iron shielding.

\begin{figure}[tbp]
	\centering
	\includegraphics[width=0.45\textwidth]{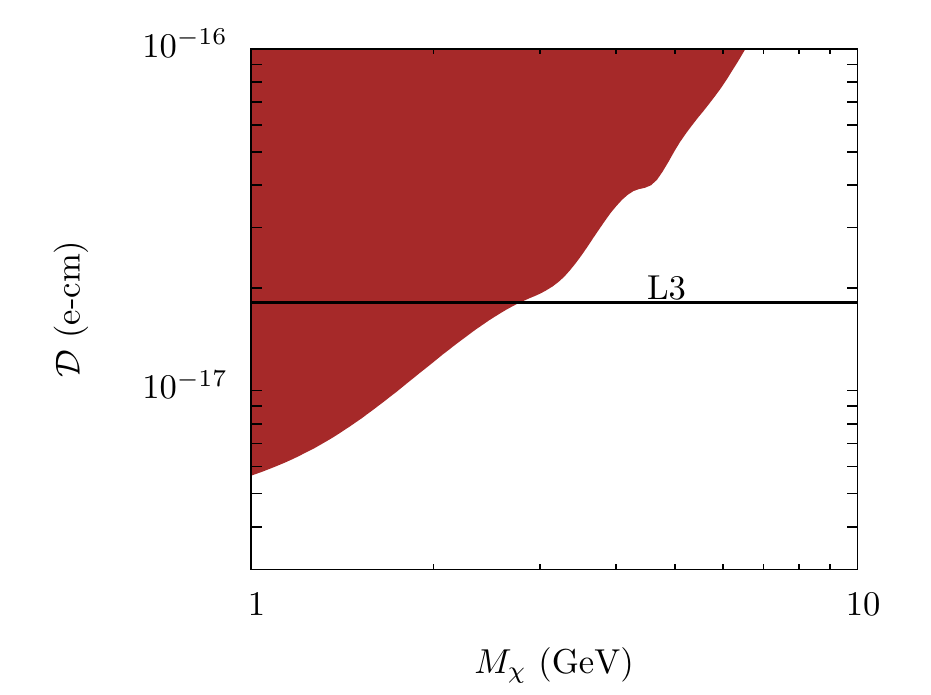}
	\includegraphics[width=0.45\textwidth]{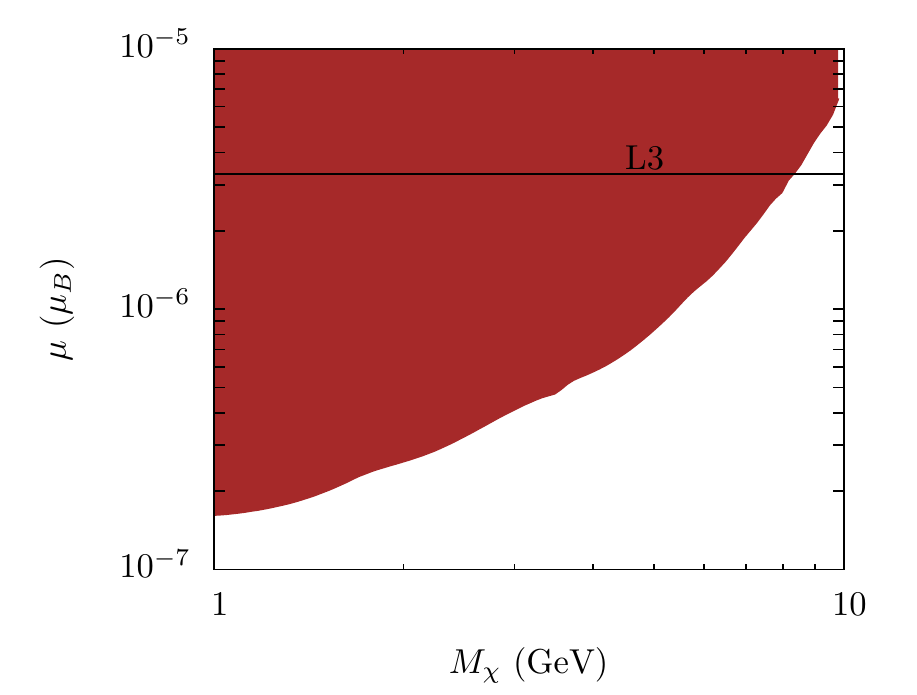}
	\caption{Left panel shows allowed parameter space (unshaded region) for electric
		dipole moment of DM in the $m_\chi-\mathcal{D}$ plane while right panel
		shows allowed parameter space for magnetic dipole moment of DM in the
		$m_\chi-\mu$ plane.}
	\label{fig:ddm}
\end{figure}

\section{Constraints on Dipolar DM from beam dump (E613) and other experiments}

From the number of events calculated using eq.(\ref{ndet}), we constrain the magnetic and
electric dipole moments of DM in the DM mass range of $1-10$ GeV.  We use the
interpretation in \cite{golowich} also used in \cite{tait} of the experimental data in
\cite{romanowski,duffy}.  We allow only those values of $m_\chi$ and $\mu,\mathcal{D}$ for
which the number of expected events is $<180$.  The results for electric and magnetic
dipole interactions of DM are shown in Fig.~\ref{fig:ddm}.  We have also added the bound
from L3 experiment at LEP on the magnetic moment of DM \cite{l3}.  The bounds on electric
and magnetic dipole moments of DM from the analysis of L3 experiment are $<1.8\times
10^{-17}$ e-cm or equivalently $<3.3\times 10^{-8}\mu_B$.  We see from the figure that for
electric dipole interaction of DM the allowed DM mass ranges between $1-3$ GeV, however
for the magnetic dipole case the DM mass lies between $1-8$ GeV.  In addition to the bound
from L3 collaboration at LEP, recently constraints from solar physics on dipolar DM and
similar models have also been studied \cite{silk,scott}.  In \cite{silk} the bound on
magnetic dipole moment of DM from helioseismological data is estimated to be $1.6\times
10^{-17}$ e-cm, for DM mass $<4.3$ GeV, which is quite similar to the bound from L3
collaboration.  Also for momentum and/or velocity dependent scattering of DM studied in
\cite{scott,scott-prl,scott3} which is relevant for dipolar DM model since it also has
momentum and velocity dependence, the most favorable DM mass is found to be $\sim 3$ GeV.
Thus we see that the constraints from beam dump experiments are in broad agreement with
those from solar physics.  Also the most stringent bounds from direct detection currently
come from LUX collaboration \cite{lux}, however this bound is considerably weak for low
mass DM particularly in the $1-10$ GeV range.  As a result this bound is inconsequential
for the results presented here.

\section{Conclusions}

We study the Dipolar DM model in the context of E613 beam dump experiment.  We see that
the constraint on electric and magnetic dipole interactions of DM from E613 experiment for
light DM in the $1-10$ GeV mass range is quite stringent.  It is restricted in that mass
range mainly by the bound on magnetic dipole moment from the L3 collaboration at
$<1.8\times 10^{-17}$ e-cm or $3.3\times 10^{-8}\mu_B$.  Bounds from solar physics data
are also broadly compatible with this mass range and dipole moment.  Thus the dipolar
model of DM offers an alternative that is compatible with constraints from wide ranging
experiments like beam dump or fixed target experiments as well as helioseismological data.
In addition the low mass range of DM enables it to be compatible with the most stringent
direct detection bounds from LUX\cite{lux}.

\end{document}